# Uncovering Multiple Intrinsic Chiral Phases in $(PEA)_2PbI_4$ halide Perovskites

*Shahar Zuri*[1], *Arthur Shapiro*[1], *Leeor Kronik*,[2] *and Efrat Lifshitz*[*1]

[1]Schulich faculty of Chemistry, Solid State Institute, Russell Berrie Nanotechnology Institute, Helen Diller Quantum Information Center and the Grand Technion Energy Program, Technion, Haifa 3200003, Israel

[2]Department of Molecular Chemistry and Materials Science, Weizmann Institute of Science, Rehovoth 76100, Israel

*Corresponding author. Email: ssefrat@techunix.technion.ac.il

Keywords: perovskites, 2D materials, anharmonicity, chiral crystals, optical materials.

### Abstract

Two-dimensional (2D) halide perovskites offer a unique platform to investigate the ground-state of materials possessing significant anharmonicity. In contrast to three-dimensional perovskites, their 2D counterparts offer substantially fewer degrees of freedom, resulting in multiple well-defined crystal structures. In this work, we thoroughly investigate the anharmonic ground-state of the benchmark $(PEA)_2PbI_4$ compound, using complementary information from low-temperature x-ray diffraction (XRD) and photoluminescence spectroscopy, supported by density functional theory (DFT) calculations. We extrapolate four crystallographic configurations from the low-temperature XRD. These configurations imply that the ground-state has an intrinsic disorder stemming from two coexisting chiral sub-lattices, each with a bi-oriented organic spacer molecule. We further show evidence that these chiral structures form unevenly populated ground-states, portraying uneven anharmonicity, where the state population may be tuned by surface effects. Our results uncover a disordered ground-state that may induce intrinsic grain boundaries, which cannot be ignored in practical applications.

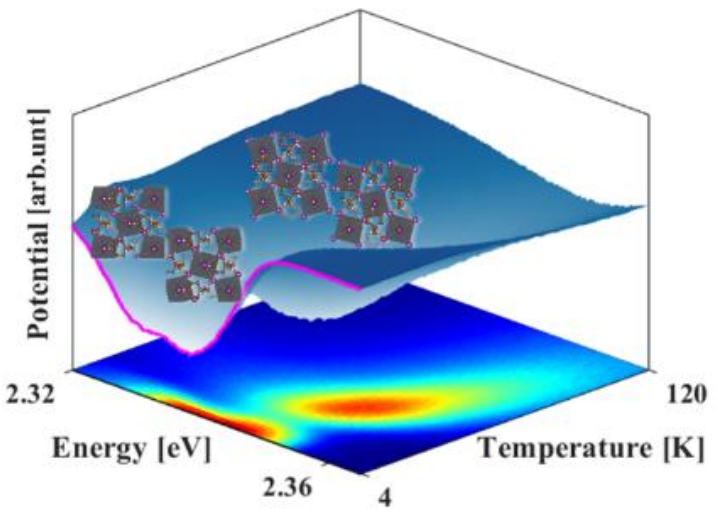

## Introduction

Hybrid perovskites (HPs) have garnered considerable attention in recent years, owing to an intriguing combination of physical properties, from tunable optical bandgaps, through high absorption coefficients to long carrier diffusion lengths, leading to high energy conversion efficiencies when used in photovoltaic cells[1]. Combining these properties with low-cost fabrication, HPs have become a prominent candidate for solar-cell and other optoelectronic applications[2].

The common three-dimensional (3D) HP crystal has a general chemical formula $ABX_3$, where metal ions (B = e.g., Pb, Sn) and halides (X = I, Cl, Br) create a network of corner-sharing octahedra. At the same time, voids within the complex are filled by organic or inorganic counter ions (A = e.g., methylammonium, $Cs^+$). This structure can also crystallize as a two-dimensional (2D) stack with a stoichiometry of $R_2A_{n-1}B_nX_{3n+1}$, where *n* inorganic octahedral slabs are held together by weak van der Waals interactions, mediated by bridging intercalated molecules (R = e.g., butylamine - BA, Phenylethylamine - PEA). Usefully, the 2D structure dramatically increases the HP moisture stability[3,4], while maintaining the properties mentioned above[5,6]. Both 3D and 2D structures possess a variety of temperature-induced phases, starting from a highly symmetric cubic phase at high temperatures to low symmetry monoclinic/triclinic phases at low temperatures. These phases exhibit different spatial arrangements of the $BX_6$ octahedra, which in turn dictate the electronic and optical properties of HPs[4,7,8].

Conventional semiconductors are characterized by a single thermodynamically stable ground-state, with an energy landscape approximated by a harmonic potential of atomic positions (Figure 1a, blue). In contrast, HPs tend to exhibit significant anharmonicity[9] and portray continuous fluctuations along a shallow potential (Figure 1a, orange) or between specific potential minima (Figure 1a, green). Furthermore, the mentioned anharmonicity may be perturbed such that an uneven atomic configuration probability (Figure 1a, violet) is found. The anharmonic nature has immense implications for HP properties and likely contributes to perplexing seeming contradictions, e.g., long carrier lifetimes but a limited diffusion length, or low Urbach energies in a disordered crystal[9–11].

The occurrence of anharmonicity in the benchmark 3D $CsPbX_3$ and $MAPbX_3$ HPs was deduced from a combination of low-frequency Raman measurements and first principles molecular dynamics, revealing significant dynamic disorder[12]. Other studies suggested that the metastable $CsPbI_3$ cubic (black) phase at high temperatures can maintain its black phase at low temperatures by anharmonic vibrations[13,14]. The ion oscillation time in that black phase was estimated to be ~ 400 fs at room temperature, exposing fast dynamics in the exchange of atomic arrangements[14]. The fast anharmonic oscillation has been related to a dynamic spin-orbit coupling (SOC) Rashba effect, which leads to a dynamic symmetry breaking that lifts the electronic degeneracy in *k*-space[15]. Significantly shortened phonon lifetimes, stemming from the structural anharmonicity, explain the relatively low thermal conductivity found in HPs[16]. The short-range dynamic distortion of transversal halide displacements explains the sharp absorption edge in $CsPbX_3$ compounds, useful for a photovoltaic application[17,18].

Anharmonic effects are also significant in 2D-HPs, where polarization-orientation Raman spectroscopy was used to show that strong anharmonic octahedral tilting plays a major role in the low-temperature phase formation of $(BA)_2PbI_4$[19]. Still, below an appropriate activation temperature, vibrations are reduced substantially and thereby locking the 2D-HPs in a frozen glass-like state. Several investigations showed that the frozen inorganic moiety imposes changes in the magnitude of the exciton diffusion, binding energy, and effective mass[19–21]. However, many open questions remain unanswered regarding the structure of this frozen disordered environment and its effect on electro-optical properties.

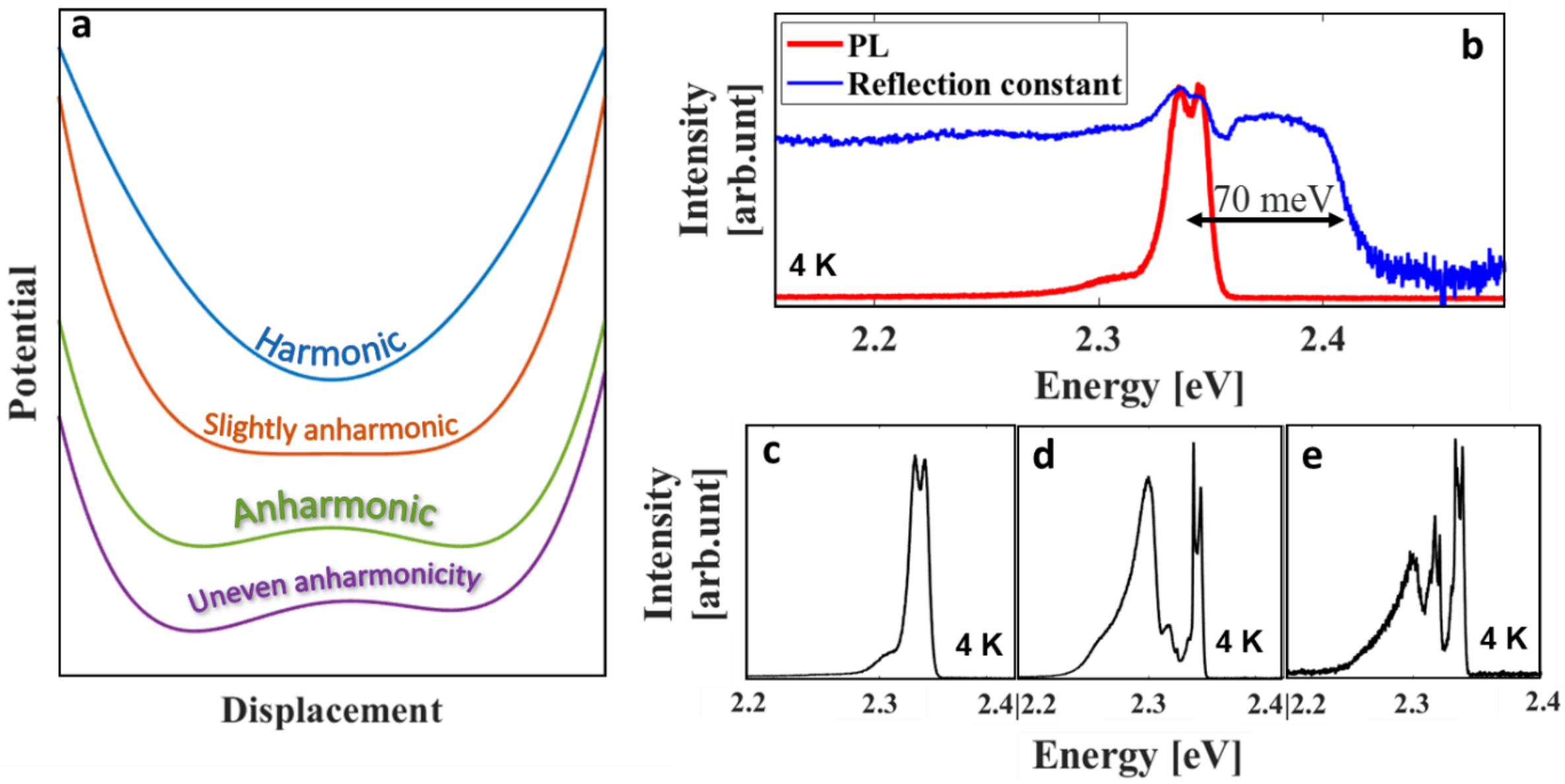

**Figure 1. Anharmonicity and its effect on the emission spectrum of a single $(PEA)_2PbI_4$ crystal. a**, Schematic description of four general types of crystals (harmonic, slightly anharmonic, anharmonic, unevenly anharmonic). **b**, Representative PL and Reflection spectra of the mostly excitonic emission at 4 K. **c-e**, PL spectra from three different places in the inhomogeneous single-crystal at 4 K.

This work focuses on the anharmonic ground-state of $(PEA)_2PbI_4$ using complementary information extracted from low-temperature x-ray diffraction (XRD) and photoluminescence (PL) spectroscopy, supported by DFT calculations. We show conclusively that the nominally single crystal $(PEA)_2PbI_4$ has a low-temperature multi-configurational crystal structure. These phase configurations imply that the ground-state of $(PEA)_2PbI_4$ has an intrinsic disorder, leading to the simultaneous formation of two coexisting chiral sub-lattices, each with a dual orientation of the organic spacer molecule.

## Results and discussion

The present study employed confocal PL spectroscopy to examine disorder effects on the optical properties of a nominally single crystal of $(PEA)_2PbI_4$. The room temperature XRD of the mentioned single crystal is given in Figure S1 of the supplementary information (SI). Figure 1b depicts typical reflectance and micro-PL spectra of the examined crystal, recorded at 4 K. This exposes dominant dual exciton bands, red shifted from the band-edge with a binding energy of ~70 meV, consistent with previous studies[22–24]. Representative PL spectra recorded at 4K and monitored at various surface spots (each about 0.5 μm in size) are shown in Figure 1c-e. These spectra show that the near band-edge dual components remain unchanged, while other spectral features reflect a large variability. Previous studies assigned the low energy features to phonon replica or to trapped-state recombination[25,26]. The current study refers to the electronic changes close to the band-edge energy. Therefore, the nature of trapped carriers will be beyond the scope of this study and is not explored further. Power-dependent measurements (SI, Figure S3) show a linear dependence of the dual-component intensities on the excitation power, excluding the formation of a biexciton. Thus, we conclude that the observed dual emission originates from a different physical phenomenon, possibly related to structural inhomogeneities. To settle this inconsistency, single-crystal XRD was used to determine the atomic positions of a $(PEA)_2PbI_4$ single crystal. Figure 2a shows the atomic structure deduced from a XRD measurements at 100 K, where the crystal is expected to be in a frozen state[27]. The measurement uncovered a triclinic polymorph crystal structure with a p-1 space group. Polymorphism was previously reported in both 3D-[28] and 2D-HPs[29], manifesting as the dual formation of monoclinic and triclinic structures. Furthermore, a disordered polymorphous cubic phase in 3D-perovskites was invoked to explain deviations from experiments, e.g., smaller band gaps and inconsistent dielectric constants[30]. Here, we show a new type of polymorphism, where four sub-configurations of $(PEA)_2PbI_4$ occupy the same triclinic unit-cell. The observed structure varies by tilting the $PbI_6$ octahedral cages and orthogonally rotating the organic PEA molecules, as emphasized by the lines in Figure 2a.

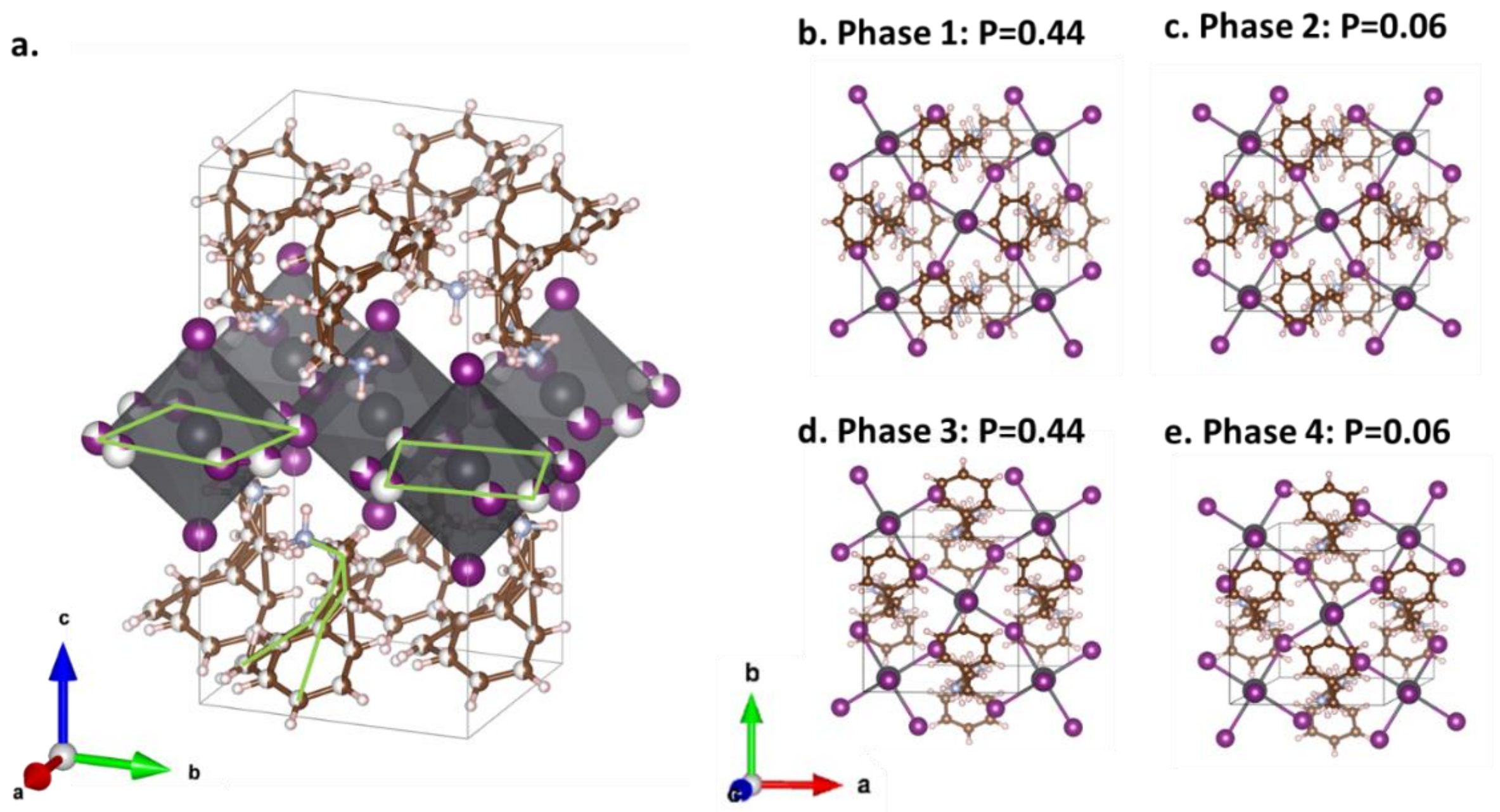

**Figure 2. Single-crystal XRD observations. a**, Structure determined from single-crystal XRD measurement at 100 K, showing a polymorphous triclinic unit-cell of $(PEA)_2PbI_4$ (see SI, Tables S1-8, for raw XRD data). Different atoms are represented by a false full-color. Semi-colored atoms denote partially occupied atom. The green lines emphasize the different orientations of the inorganic octahedra and the organic PEA molecules. **b-e**, a detailed description of the 4 configurations, as isolated from the crystallographic results into distinct phases, 1-4. *P*, extracted from the partial occupancy parameter, denotes the occurrence probability for each phase.

To find the stable ground-state configuration, polymorph components were separated into four frozen sub-configurations, as shown in Figure 2b-e. The XRD data show that the four configurations arise from the combination of two key structural differences: phase 1 (and 2) differs from phase 3 (and 4) by an orthogonal rotation of the organic PEA molecule. More importantly, the inorganic parts of phase 1 (and 3) and phase 2 (and 4) are *chiral pairs* of each other, resulting from a tilt in the rigid $PbI_6$ octahedra perpendicular to the $c^*$ crystallographic axis. These two degrees of freedom, namely the chirality of the inorganic framework and the orthogonal rotation of the organic spacer molecule, account for the four possible local energy minima. We note that induced structural chirality was previously demonstrated in 2D-HPs by inserting chiral organic R cations[31]. Our results show a case where structural chirality is an intrinsic property of these crystals, regardless of chirality stemming from the spacer molecules. Other configurations were excluded by chemical implausibility. Thus, we determine this crystal to be in the anharmonic regime, in the sense of Figure 1a (purple curve). Each configuration is roughly assigned with an occurrence probability $P$, extracted from the partial occupancy parameter of the single crystal XRD measurement (Table. S1). $P$ is an important parameter, reflecting the occurrence of each frozen configuration down to 100 K and below.

To gain further insight into the ground-state configuration, we performed DFT calculations for all four phases. We chose fully-relativistic pseudopotentials to account for the strong SOC in Pb and I[32–34] and explicitly added dispersion correction terms[35,36]. The ground-state structures and response properties are routinely determined from DFT, based on room-temperature XRD coordinates that average over the energy landscape. Soft materials like HPs challenge the usual approach and may be misinterpreted as a global minimum energy structure. 2D-HPs may amplify this difficulty through the highly distorted nature of both organic and inorganic moieties. We therefore considered all four configurations as a ground-state candidate and relaxed all of them starting from the pertinent experimental coordinates.

**Table 1. Computed DFT variables for the proposed four phase configurations relative to phase 1. $\Delta E_{tot}$ is the total DFT energy, $|\Delta T|$ is the equivalent thermal energy, and $\Delta E_g^{HSE}$ is the relative band-gap calculated by DFT+HSE. See the methods section for details.**

| Phase | $\Delta E_{tot}$ [meV/atom] | $\|\Delta T\|$ [K] | $\Delta E_g^{HSE}$ [meV] |
|---|---|---|---|
| 1 | 0 | 0 | 0 |
| 2 | -0.1 | 2.4 | 30.3 |
| 3 | 1.8 | 31.8 | 35.8 |
| 4 | 1.9 | 32.2 | 16.6 |

Interestingly, the relaxed structures showed small total energy variations, as summarized in Table 1. Our DFT calculations find a very slight reduction of the total energy of phase 2 compared to the reference phase 1, with a difference of 0.1 meV. Such a minor difference is essentially within the error range of the DFT calculation, suggesting that phases 1 and 2 are energetically equivalent. Phases 3 and 4 show higher total energies compared to phases 1 and 2, with a similar energy difference between them.

Thermodynamically, the most stable configuration should have the lowest formation energy at the low-temperature limit, as described by[28,37]:

$$\Delta H_f^i = E^i[(PEA)_2PbI_4] - (E[(PEA)] + E[PbI_2]), \quad (1)$$

where $E^i[(PEA)_2PbI_4]$ is the total energy, $i$ is the configuration, and $E[(PEA)] + E[PbI_2]$ is the total energy of the constituents. Because the formation energy of the four configurations differs only by the first term of Eq. (1), a relative comparison between the configurations, based on DFT total energies, suffices. It is then convenient to describe the formation energy difference of the phases in terms of the equipartition thermal energy, $3/2k_bT$, such that $|\Delta T|$ in Table 1 represents an approximate stable temperature for each configuration relative to the 0 K ground-state (i.e., phases 1 and 2). This interpretation suggests that phases 3 and 4 form stable configurations at around 30 K, manifesting as a configuration transition brought about by the collective rotation of the organic molecules.

To further examine this finding, we performed temperature-dependent PL (TPL) measurements. Figure 3a shows that the low-temperature PL at 4 K originates from dual excitonic emissions separated by a 10 meV. Additional PL spectra at different temperatures are provided in Figure S4. Because phases 1 and 2 were found to form a dual ground state, both should emit at the same energy. However, small orientation changes between the amine group of the organic molecule and the inorganic frameworks slightly break the symmetry between the two phases, lifting the degeneracy between them (Figure S2)[19]. Interestingly, our confocal apparatus allowed us to probe different slabs of the crystal by raising the focal plane from the bulk to the surface. We found that the ground-state emission associated with phases 1 and 2 corresponds to different recombination processes from the surface and bulk regions (see labels in Figure 3b), with an energy difference of 10 meV. A similar trend was noted recently via a two-photon absorbance method, finding an energy gap of around 80 meV between a surface and a bulk emission band in the room-temperature PL spectrum[38]. There, dual emission was explained as surface trap recombination. However, owing to the lower energy difference and the lack of power dependence associated with trap-state emissions, we suggest that the dual emission here originates from the two coexisting phases.

To further explore the dual emission, we considered the intensity partial contribution (IPC), defined as $(I_s\text{-}I_b)/(I_s\text{+}I_b)$, where $I_b$ and $I_s$ are the dominant bulk and surface integrated intensities, respectively. An exciton can statistically form in both configurations, recombining into different local potentials. The recombination event is an observable quantum mechanical property. So, by tracing the excitonic recombination in a PL spectrum, one can deduce the origin of the configuration. We therefore consider the observable IPC as a proxy for the occurrence parameter $P$, i.e., the IPC can be roughly calculated as $(P_2\text{-}P_1)/(P_2\text{+}P_1)$. Figure 3c shows the IPC as a function of the focal depth. The measured IPC plateaus at ~25% when the focal point is lowest (i.e., inner depth), where the computed value based on *P is* 24%. We note that the XRD-extracted value of *P* mostly indicates the inner emission as the surface layers are negligible with respect to the bulk crystal. The rotation of the organic moieties below 20 K causes phases 3 and 4 to readjust as phases1 and 2, respectively. However, the ratio between phases 1 and 2 is maintained, thus the IPC derived at 100 K persist down to 4 K. Remarkably, when moving to the surface (i.e., outer depth), we observe a linear increase of the measured IPC until reaching a plateau at the surface with an inverted picture of IPC at around 75%. This suggests that the surface changes the formation of phases 1 and 2 to the point of completely flipping the phase population. Therefore, we find that the nature of near-surface and bulk excitons arises from an intrinsic property of the crystal derived from a configurational equivalence of the crystal structure. Furthermore, the two ground-state configurations form intra-layer domains, where one of the phases is further stabilized near the crystal surface. This new observation directs us to view the ground-state configurations of $(PEA)_2PbI_4$ as real-space manifestations of uneven anharmonicity, in the sense described above.

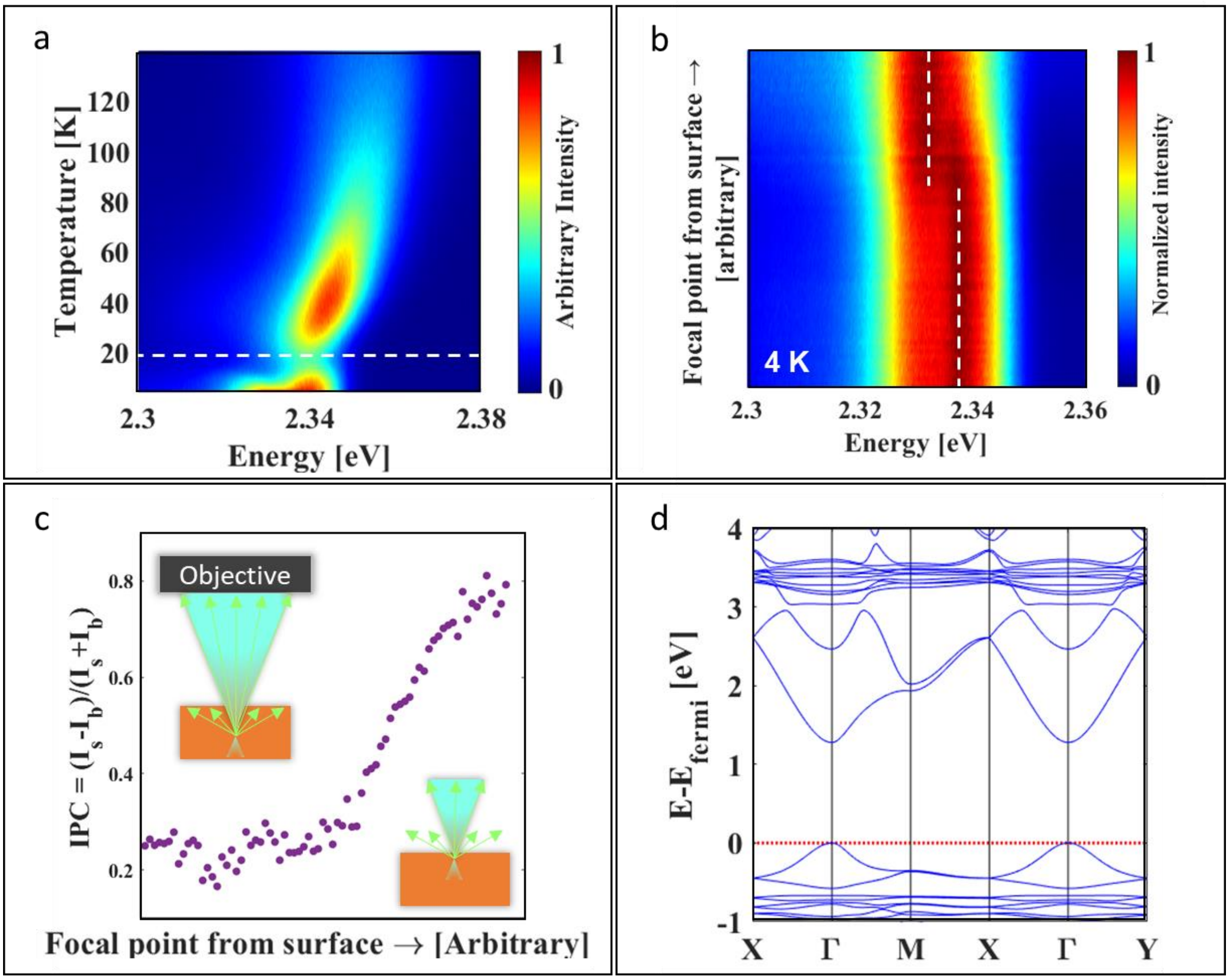

**Figure 3. Opto-electronic and DFT results for the band edge of single-crystal $(PEA)_2PbI_4$. a**, Temperature-dependent PL colormap of the excitonic region. The dashed line at 20 K depicts a phase transition between phases 1 and 2 to phases 3 and 4. **b**, PL color map as a function of focal point distance from the surface and the energy of the dual emission bands (see dashed lines) at 4 K. **c**, Plot of the IPC value as a function of the focal point distance from the surface. Two extreme schematic

illustrations of the confocal arrangement are described: Left) The objective lens focal point is closer to the bulk crystal, thus collecting light from a deeper part of the crystal. Right) The objective lens focal point is at the surface of the bulk crystal, thus collecting light from the outer surface. **d**, DFT band structure of phase 1.

In the TPL measurements, we also observe an abrupt shift in the PL trend at 20 K accompanied by a drastic intensity decline, as marked by the white dashed line in Figure 3a. Similar behavior in the TPL was observed previously in PbS[39] and CdSe[40] quantum dots and explained by a collective rearrangement of surface organic ligands. In our case, the calculated phase transition between phase 1 (and 2) to phase 3 (and 4) derives from the collective rotation of the organic PEA molecules, which agrees well with the DFT prediction for a phase transition above 20 K. Note that to determine precisely whether all four configurations coexist above 20 K, entropic terms must be considered, which lies beyond the scope of this work. Furthermore, similar low temperature phase transitions were observed in previous works on $(PEA)_2SnI_4$[41] and $(PEA)_2PbX_4$[42,43], whereas samples with narrower cross-section molecules like BA did not show this behavior[44]. Hence, we deem this structural polymorphism to be a collective property shared across many more 2D-HPs with wide intercalated molecules.

To further explore the electronic structure of the four configurations, we calculated the electronic band structure of phases 1-4. A representative band structure (of phase 1) is shown in Figure 3d and Figure S5 in the SI shows that all four configurations have a similar band structure, devoid of a Rashba-induced spin splitting. We note that this is an unsurprising result because all configurations share a centrosymmetric p -1 triclinic structure, where the electronic band-edge states derive from the $PbI_6$ octahedra. Thus, phases 1 (2) and 3 (4) are electronically equivalent. Moreover, as phases 1 and 2 differ only in a chiral fashion with an inversion center, their overall band-structure should remain the same. However, we observe small changes in the calculated bandgap, as summarized in Table 1. The absolute DFT bandgap values are well-expected to deviate from the experimental bandgap energies[45]. Yet, the small changes in the crystal structure allow a relative comparison between the four configurations. Our results show that the DFT bandgap difference between phases 1 and 2 is ~30 meV, of the order of the observed low-temperature dual emission difference of 10 meV (Figure 3a-b). Thus, we can infer that phases 1 and 2 correlate with the bulk stabilized and surface stabilized emission processes, respectively.

In this work, we showed that the four-configuration model consistently explained the TPL spectra, unveiling that the fine-structure excitonic emission comes from a pair of stabilized configurations. Furthermore, the observed stable coexisting interlayer configurations and the extrapolated 12:88 probability of forming each domain portrays a new type of unevenly distributed anharmonic ground state. Based on our results, we predict a large number of disordered domain boundaries in $(PEA)_2PbI_4$, with a local non-centrosymmetric structure. We propose that such an intrinsic granular environment should be explored further as the source of spatial inhomogeneities that induce location-dependent traps and phonon coupling strengths. We further speculate that this intrinsic, robust creation of grain boundaries can induce a Rashba effect, even in the nominally centrosymmetric structure of $PEA_2PbI_4$, an issue that will be elaborated elsewhere. At high temperature, more configurations may develop from this frozen multi-configurational environment, arising from the highly anharmonic nature of these crystals. These configurations can ultimately affect charge transport and spin coherence and therefore cannot be neglected when addressing the optoelectronic properties of these materials.

## Methods

### Synthesis

$(PEA)_2PbI_4$ single crystals were synthesized via the anti-solvent recrystallization method. First, HI (5 mL), $H_3PO_2$ (0.1 mL), and $PbI_2$ (0.461 gr) precursors were dissolved in acetone (3 mL) at 70 °C until a clear yellow solution was obtained. Afterwards, PEA (0.2514 mL) precursor was inserted with raising the temperature to 80 °C. The reaction was maintained for 15 minutes and then cooled using a water bath to room temperatures where $(PEA)_2PbI_4$ flakes formed. The crystals obtained were washed thoroughly by vacuum filtration using diethyl ether as a washing solvent and dried under vacuum (inside desiccator). To form single crystals, an additional step was added. The crystals were dissolved in acetonitrile and heated a on a plate with a tightly closed cap till a saturated solution was obtained. The vile containing the dissolved crystals was placed in a toluene bath. The vile was left open in the closed toluene bath for 3 days until single crystals formed at the liquid-gas interface. Lastly, the crystals were washed by diethyl ether and then dried inside a desiccator.

Optical confocal spectroscopy

The optical properties of the 2D-HPs were measured on a fiber based confocal probe**.** The sample was placed on a glass substrate, mounted on an XYZ attocube piezo-electric stage, allowing for controlled movement over the substrate. The optical probe was immersed in a cryogenic system (attoDRY1000 closed cycle cryostat) with an objective lens of NA=0.65. A ceramic heater on the sample stage allowed for a TPL measurements, where the temperature of the sample was increased gradually from 4-290 K. The microscope used a 488nm laser (iBeam, TOPTICA) as an excitation source and a 495nm long-pass dichroic mirror (SEMROCK, FF495-Di03-25x36) where the photo-emission were collected through optical fibers. For PL detection a monochromator (Acton SP2358) with 1200gr/mm blz 500nm grating and EM-CCD (FI Newton Andor) were used. To investigate the bulk and surface emission, the Z piezo was tuned to "fine" step mode, roughly obtaining step length of tens to hundred nanometers. Reflection measurement used a reference Si substrate and a tungsten white-light source (Thorlabs).

DFT simulation

First principles calculations were carried out using the PBE (Perdew–Burke–Ernzerhof) generalized gradient exchange correlational functional[46] using pseudopotentials to treat core electron effects, as implemented in the open-source Density functional theory (DFT) package Quantum Espresso[47]. All calculations were carried out with planewave basis using a plane-wave kinetic energy cutoff of 90 Ry. Dispersion interactions were accounted for using the Grimme-D3 method[48]. For a more accurate description of the bandgap, a modified Heyd-Scuseria-Ernzerhof functional[49], with a short-range exchange fraction of 42%[50], was used.

## Data availability

The data supporting the findings of this study are available within the paper and Supplementary Information. Source data are provided with this paper.

## Acknowledgements

This work was supported by the Israel Science Foundation (ISF, project no. 2528/19) and by the Minerva Centre for Self-Repairing Systems for Energy & Sustainability. The authors thank Dr. Natalia Friedman (Technion) for carrying the single-crystal XRD measurements and data refinement, Dr. Maria Koifman (Technion) for measuring the room-temperature XRD, and Guy Ohad (Weizmann Institute of Science) for useful computational advice. L.K. thanks the Mintzi and Aryeh Katzman Professorial Chair, the Helen and Martin Kimmel Award for Innovative Investigation, and a research grant from the Perlman-Epstein Family C-AIM Impact Fund for Survivability and Sustainability. This research used computational resources provided by the Molecular Foundry of the National Energy Research Scientific Computing Center, a DOE Office of Science User Facility supported by the Office of Science, Office of Basic Energy Sciences, of the U.S. Department of Energy under Contract No. DE-AC02-05CH11231.

*Supplementary information*

# Uncovering Multiple Intrinsic Chiral Phases in $(PEA)_2PbI_4$ halide Perovskites

*Shahar Zuri[1], Arthur Shapiro[1], Leeor Kronik,[2] and Efrat Lifshitz*[1]*

[1]Schulich faculty of Chemistry, Solid State Institute, Russell Berrie Nanotechnology Institute, and the Helen Diller Quantum Information Center, Technion, Haifa, Israel

[2] Department of Molecular Chemistry and Materials Science, Weizmann Institute of Science, Rehovoth 76100, Israel

*Corresponding author

Email: ssefrat@techunix.technion.ac.il

## X-ray diffraction (XRD)

XRD measurements of a $(PEA)_2PbI_4$ single crystal (PEA = phenylethylamine) - were carried out at two different temperatures. Figure S1 displays an XRD spectrum recorded at room temperature with the crystal laid along the stacking crystallographic axis. This spectrum shows a typical single crystal pattern comprised of the high-order diffractions along the (0 0 l) direction.

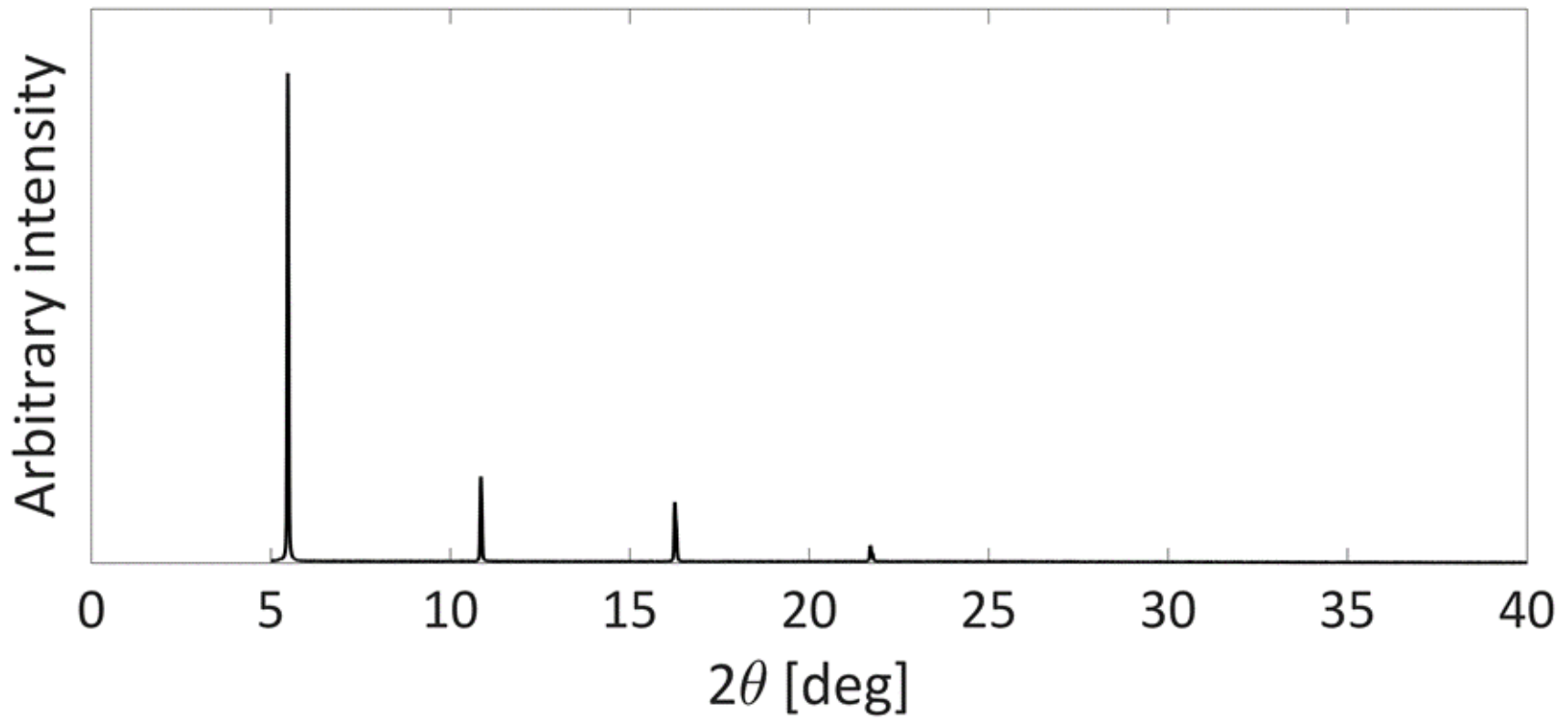

Figure S1: XRD of single crystal $(PEA)_2PbI_4$ taken at room temperature

The XRD measurement at 100 K supplied detailed crystallographic parameters (e.g., atomic position, crystallographic factors), given in Tables S1-8 below.

Table S 1: Crystal data and structure refinement for $(PEA)_2PbI_4$ single crystal

| | |
|---|---|
| Identification code | Efrat4b |
| Empirical formula | $C_{16}H_{24}I_4N_2Pb$ |
| Formula weight | 959.16 |
| Temperature/K | 100.15 |
| Crystal system | triclinic |
| Space group | P-1 |
| a/Å | 8.7029(12) |
| b/Å | 8.7042(12) |
| c/Å | 16.483(2) |
| α/° | 100.307(2) |
| β/° | 94.724(3) |
| γ/° | 90.388(4) |
| Volume/Å$^3$ | 1224.0(3) |
| Z | 2 |
| $\rho_{calc}$g/cm$^3$ | 2.602 |
| μ/mm$^{-1}$ | 11.937 |
| F(000) | 856.0 |
| Crystal size/mm$^3$ | 0.18 × 0.15 × 0.06 |
| Radiation | MoKα (λ = 0.71073) |
| 2Θ range for data collection/° | 4.698 to 50.26 |
| Index ranges | $-10 \le h \le 10$, $-10 \le k \le 10$, $-19 \le l \le 19$ |
| Reflections collected | 10183 |
| Independent reflections | 4325 [$R_{int}$ = 0.0486, $R_{sigma}$ = 0.0720] |
| Data/restraints/parameters | 4325/454/325 |
| Goodness-of-fit on $F^2$ | 1.084 |
| Final R indexes [I>=2σ (I)] | $R_1$ = 0.0468, $wR_2$ = 0.1155 |
| Final R indexes [all data] | $R_1$ = 0.0644, $wR_2$ = 0.1240 |
| Largest diff. peak/hole / e $Å^{-3}$ | 2.56/-1.73 |

Table S 2: Fractional Atomic Coordinates (×$10^4$) and Equivalent Isotropic Displacement Parameters ($Å^2$×$10^3$) for $(PEA)_2PbI_4$. U(eq) is defined as 1/3 of of the trace of the orthogonalized $U_{IJ}$ tensor.

| **Atom** | ***x*** | ***y*** | ***z*** | **U(eq)** |
|---|---|---|---|---|
| Pb | 0 | 5000 | 5000 | 15.68(19) |
| $Pb^{(1)}$ | 5000 | 10000 | 5000 | 15.63(19) |
| $I^{(1)}$ | 5183.1(9) | 10655.2(10) | 6981.6(5) | 30.3(2) |
| $I^{(2)}$ | 1898(10) | 8117(10) | 5019(5) | 54(3) |
| $I^{(2A)}$ | 3115.1(9) | 6873.7(8) | 4980.3(6) | 19.3(3) |
| $I^{(3)}$ | 3120(10) | 3111(11) | 5000(5) | 50(3) |
| $I^{(3A)}$ | 1897.8(9) | 1900.1(8) | 5000.2(6) | 19.2(3) |
| $I^{(4)}$ | 331.7(9) | 5809.9(9) | 6982.9(5) | 29.7(2) |
| $N^{(1)}$ | 4620(30) | 3260(30) | 2100(16) | 45(3) |
| $N^{(1A)}$ | 5300(30) | 3970(30) | 2119(16) | 46(3) |
| $C^{(1)}$ | 4490(30) | 4540(30) | 1580(17) | 35(4) |
| $C^{(1A)}$ | 3800(30) | 3850(30) | 1589(17) | 34(5) |
| $C^{(2)}$ | 3112(17) | 5160(17) | 1373(10) | 46(3) |
| $C^{(3)}$ | 2870(30) | 6310(30) | 900(20) | 55(6) |
| $C^{(3A)}$ | 1790(30) | 5230(30) | 880(20) | 58(7) |
| $C^{(4)}$ | 4150(40) | 6820(30) | 580(20) | 57(5) |
| $C^{(4A)}$ | 1000(40) | 3840(40) | 570(20) | 73(9) |
| $C^{(5)}$ | 5500(40) | 6330(30) | 750(20) | 57(5) |
| $C^{(5A)}$ | 1540(30) | 2560(40) | 740(20) | 61(7) |
| $C^{(6)}$ | 5770(30) | 5080(30) | 1250(18) | 41(5) |
| $C^{(6A)}$ | 3010(30) | 2450(30) | 1250(19) | 46(6) |
| $C^{(7)}$ | 4340(40) | 4000(40) | 3070(20) | 46(3) |
| $C^{(7A)}$ | 5030(40) | 4690(40) | 3020(20) | 45(3) |
| $C^{(8)}$ | 5530(30) | 5210(30) | 3380(20) | 46(3) |
| $C^{(8A)}$ | 3970(40) | 3630(40) | 3380(20) | 45(3) |
| $N^{(2)}$ | -570(30) | -1690(30) | 2126(16) | 49(4) |
| $N^{(2A)}$ | 280(30) | -920(30) | 2132(15) | 46(5) |
| $C^{(9)}$ | -600(40) | -510(30) | 1567(18) | 36(4) |
| $C^{(9A)}$ | -1230(30) | -1160(40) | 1554(19) | 36(5) |
| $C^{(10)}$ | -2006(19) | 126(19) | 1307(10) | 50(3) |
| $C^{(11)}$ | -1810(40) | 1280(40) | 818(19) | 58(6) |

| Atom | *x* | *y* | *z* | U(eq) |
|---|---|---|---|---|
| C$^{(11A)}$ | -3370(40) | -200(40) | 849(19) | 54(6) |
| C$^{(12)}$ | -400(40) | 1830(40) | 580(20) | 65(7) |
| C$^{(12A)}$ | -3960(40) | -1750(40) | 570(20) | 55(6) |
| C$^{(13)}$ | 890(40) | 1150(40) | 870(20) | 55(6) |
| C$^{(13A)}$ | -3160(40) | -2990(50) | 850(20) | 55(6) |
| C$^{(14)}$ | 800(30) | 60(30) | 1356(18) | 43(5) |
| C$^{(14A)}$ | -1860(30) | -2640(30) | 1328(18) | 42(6) |
| C$^{(15)}$ | -37(17) | -977(15) | 3029(9) | 41(3) |
| C$^{(16)}$ | -1041(12) | 232(12) | 3369(8) | 22(2) |

Table S 3: Anisotropic Displacement Parameters (Å$^2$×10$^3$) for $(PEA)_2PbI_4$. The Anisotropic displacement factor exponent takes the form: $-2\pi^2[h^2a^*2U_{11}+2hka^*b^*U_{12}+\ldots]$.

| Atom | $U_{11}$ | $U_{22}$ | $U_{33}$ | $U_{23}$ | $U_{13}$ | $U_{12}$ |
|---|---|---|---|---|---|---|
| Pb | 12.4(3) | 11.4(3) | 24.1(4) | 5.4(2) | 1.6(2) | -1.1(2) |
| Pb$^{(1)}$ | 12.7(3) | 11.0(3) | 24.3(4) | 6.1(2) | 1.7(2) | -0.1(2) |
| I$^{(1)}$ | 25.8(5) | 40.5(5) | 24.1(5) | 5.3(4) | 1.0(3) | -3.9(3) |
| I$^{(2)}$ | 68(6) | 58(5) | 34(5) | 5(4) | 2(4) | -33(4) |
| I$^{(2A)}$ | 13.8(4) | 11.3(4) | 33.8(6) | 5.8(3) | 4.3(3) | -5.7(3) |
| I$^{(3)}$ | 50(5) | 66(5) | 35(5) | 10(4) | 8(4) | -11(4) |
| I$^{(3A)}$ | 12.9(4) | 13.0(4) | 33.2(6) | 8.6(3) | 1.4(3) | 7.3(3) |
| I$^{(4)}$ | 41.2(5) | 24.1(4) | 24.4(5) | 6.4(3) | 1.6(4) | -3.7(3) |
| N$^{(1)}$ | 51(9) | 37(8) | 49(8) | 16(6) | -7(6) | -13(6) |
| N$^{(1A)}$ | 56(9) | 37(8) | 46(7) | 17(6) | -8(6) | -4(6) |
| C$^{(1)}$ | 36(9) | 29(10) | 41(11) | 5(8) | 5(9) | 6(8) |
| C$^{(1A)}$ | 38(11) | 36(9) | 30(10) | 11(9) | 10(8) | 7(8) |
| C$^{(2)}$ | 38(7) | 50(7) | 53(9) | 18(6) | 7(6) | 8(5) |
| C$^{(3)}$ | 64(13) | 46(14) | 57(17) | 15(10) | 0(12) | 23(11) |
| C$^{(3A)}$ | 46(14) | 73(13) | 56(17) | 13(13) | 10(11) | 44(11) |
| C$^{(4)}$ | 81(12) | 41(11) | 53(14) | 21(9) | 7(11) | 6(10) |
| C$^{(4A)}$ | 56(17) | 114(17) | 45(19) | 10(17) | -16(13) | 9(13) |
| C$^{(5)}$ | 81(12) | 41(11) | 53(14) | 21(9) | 7(11) | 6(10) |
| C$^{(5A)}$ | 53(14) | 83(13) | 41(15) | -3(13) | 0(11) | -10(12) |
| C$^{(6)}$ | 41(9) | 37(12) | 48(15) | 11(9) | 10(10) | 8(9) |
| C$^{(6A)}$ | 37(12) | 49(9) | 50(16) | 3(11) | 14(10) | -1(9) |
| C$^{(7)}$ | 56(9) | 37(8) | 46(7) | 17(6) | -8(6) | -4(6) |
| C$^{(7A)}$ | 51(9) | 37(8) | 49(8) | 16(6) | -7(6) | -13(6) |
| C$^{(8)}$ | 56(9) | 37(8) | 46(7) | 17(6) | -8(6) | -4(6) |
| C$^{(8A)}$ | 51(9) | 37(8) | 49(8) | 16(6) | -7(6) | -13(6) |

| Atom | $U_{11}$ | $U_{22}$ | $U_{33}$ | $U_{23}$ | $U_{13}$ | $U_{12}$ |
|---|---|---|---|---|---|---|
| $N^{(2)}$ | 63(13) | 31(10) | 52(10) | 2(6) | 12(9) | -3(8) |
| $N^{(2A)}$ | 38(10) | 56(12) | 40(9) | -5(9) | 6(7) | -9(8) |
| $C^{(9)}$ | 43(10) | 38(11) | 23(10) | -6(8) | 5(9) | 5(8) |
| $C^{(9A)}$ | 33(11) | 45(10) | 28(10) | -6(10) | 15(8) | 10(8) |
| $C^{(10)}$ | 56(8) | 56(8) | 34(9) | -5(6) | 11(6) | 14(6) |
| $C^{(11)}$ | 64(12) | 75(16) | 32(15) | 7(10) | -8(12) | 26(12) |
| $C^{(11A)}$ | 59(13) | 82(12) | 26(15) | 26(12) | 7(10) | 11(11) |
| $C^{(12)}$ | 89(16) | 73(16) | 37(16) | 19(12) | 9(13) | 17(12) |
| $C^{(12A)}$ | 46(11) | 85(12) | 34(12) | 12(10) | -10(9) | 3(8) |
| $C^{(13)}$ | 68(13) | 49(14) | 54(17) | 12(11) | 24(12) | 12(11) |
| $C^{(13A)}$ | 46(11) | 85(12) | 34(12) | 12(10) | -10(9) | 3(8) |
| $C^{(14)}$ | 55(10) | 42(13) | 33(14) | 2(9) | 12(11) | 1(10) |
| $C^{(14A)}$ | 37(12) | 43(9) | 33(14) | -22(10) | 1(9) | 10(9) |
| $C^{(15)}$ | 61(9) | 21(6) | 42(7) | 10(6) | 3(6) | 19(5) |
| $C^{(16)}$ | 16(6) | 22(5) | 28(7) | 7(5) | -3(5) | 0(4) |

Table S 4: Bond Lengths for $(PEA)_2PbI_4$.

| Atom | Atom | Length/Å | Atom | Atom | Length/Å |
|---|---|---|---|---|---|
| Pb | $I^{(2)1}$ | 3.161(7) | $C^{(1A)}$ | $C^{(6A)}$ | 1.40(3) |
| Pb | $I^{(2)}$ | 3.161(7) | $C^{(2)}$ | $C^{(3)}$ | 1.38(3) |
| Pb | $I^{(2A)1}$ | 3.1597(8) | $C^{(2)}$ | $C^{(3A)}$ | 1.36(3) |
| Pb | $I^{(2A)}$ | 3.1597(8) | $C^{(3)}$ | $C^{(4)}$ | 1.38(3) |
| Pb | $I^{(3)1}$ | 3.184(9) | $C^{(3A)}$ | $C^{(4A)}$ | 1.38(3) |
| Pb | $I^{(3)}$ | 3.184(9) | $C^{(4)}$ | $C^{(5)}$ | 1.28(4) |
| Pb | $I^{(3A)1}$ | 3.1731(8) | $C^{(4A)}$ | $C^{(5A)}$ | 1.28(4) |
| Pb | $I^{(3A)}$ | 3.1731(8) | $C^{(5)}$ | $C^{(6)}$ | 1.49(3) |
| Pb | $I^{(4)1}$ | 3.2063(9) | $C^{(5A)}$ | $C^{(6A)}$ | 1.49(3) |
| Pb | $I^{(4)}$ | 3.2063(9) | $C^{(7)}$ | $C^{(8)}$ | 1.46(4) |
| $Pb^{(1)}$ | $I^{(1)2}$ | 3.2042(9) | $C^{(7A)}$ | $C^{(8A)}$ | 1.53(4) |
| $Pb^{(1)}$ | $I^{(1)}$ | 3.2041(9) | $N^{(2)}$ | $C^{(9)}$ | 1.50(4) |
| $Pb^{(1)}$ | $I^{(2)2}$ | 3.154(8) | $N^{(2)}$ | $C^{(15)}$ | 1.54(3) |
| $Pb^{(1)}$ | $I^{(2)}$ | 3.154(8) | $N^{(2A)}$ | $C^{(9A)}$ | 1.55(4) |
| $Pb^{(1)}$ | $I^{(2A)}$ | 3.1617(8) | $N^{(2A)}$ | $C^{(15)}$ | 1.54(3) |
| $Pb^{(1)}$ | $I^{(2A)2}$ | 3.1617(8) | $C^{(9)}$ | $C^{(10)}$ | 1.41(3) |
| $Pb^{(1)}$ | $I^{(3)3}$ | 3.173(10) | $C^{(9)}$ | $C^{(14)}$ | 1.40(4) |
| $Pb^{(1)}$ | $I^{(3)4}$ | 3.173(10) | $C^{(9A)}$ | $C^{(10)}$ | 1.41(4) |
| $Pb^{(1)}$ | $I^{(3A)3}$ | 3.1756(8) | $C^{(9A)}$ | $C^{(14A)}$ | 1.38(4) |
| $Pb^{(1)}$ | $I^{(3A)4}$ | 3.1756(8) | $C^{(10)}$ | $C^{(11)}$ | 1.41(4) |

| Atom | Atom | Length/Å | Atom | Atom | Length/Å |
|---|---|---|---|---|---|
| $I^{(3)}$ | $Pb^{(1)5}$ | 3.173(10) | $C^{(10)}$ | $C^{(11A)}$ | 1.36(3) |
| $I^{(3A)}$ | $Pb^{(1)5}$ | 3.1756(8) | $C^{(11)}$ | $C^{(12)}$ | 1.43(5) |
| $N^{(1)}$ | $C^{(1)}$ | 1.52(3) | $C^{(11A)}$ | $C^{(12A)}$ | 1.42(5) |
| $N^{(1)}$ | $C^{(7)}$ | 1.65(4) | $C^{(12)}$ | $C^{(13)}$ | 1.37(4) |
| $N^{(1A)}$ | $C^{(1A)}$ | 1.50(3) | $C^{(12A)}$ | $C^{(13A)}$ | 1.41(5) |
| $N^{(1A)}$ | $C^{(7A)}$ | 1.53(4) | $C^{(13)}$ | $C^{(14)}$ | 1.35(4) |
| $C^{(1)}$ | $C^{(2)}$ | 1.36(3) | $C^{(13A)}$ | $C^{(14A)}$ | 1.33(4) |
| $C^{(1)}$ | $C^{(6)}$ | 1.39(3) | $C^{(15)}$ | $C^{(16)}$ | 1.438(17) |
| $C^{(1A)}$ | $C^{(2)}$ | 1.38(3) | | | |

[1]-X,1-Y,1-Z; [2]1-X,2-Y,1-Z; [3]1-X,1-Y,1-Z; [4]+X,1+Y,+Z; [5]+X,-1+Y,+Z

Table S 5: Bond Angles for $(PEA)_2PbI_4$.

| Atom | Atom | Atom | Angle/° | Atom | Atom | Atom | Angle/° |
|---|---|---|---|---|---|---|---|
| $I^{(2)}$ | Pb | $I^{(2)1}$ | 180.0 | $I^{(3)4}$ | $Pb^{(1)}$ | $I^{(1)}$ | 89.17(16) |
| $I^{(2)1}$ | Pb | $I^{(3)}$ | 90.0(2) | $I^{(3)3}$ | $Pb^{(1)}$ | $I^{(3)4}$ | 180.0 |
| $I^{(2)}$ | Pb | $I^{(3)}$ | 90.0(2) | $I^{(3A)4}$ | $Pb^{(1)}$ | $I^{(1)2}$ | 91.59(2) |
| $I^{(2)1}$ | Pb | $I^{(3)1}$ | 90.0(2) | $I^{(3A)4}$ | $Pb^{(1)}$ | $I^{(1)}$ | 88.41(2) |
| $I^{(2)}$ | Pb | $I^{(3)1}$ | 90.0(2) | $I^{(3A)3}$ | $Pb^{(1)}$ | $I^{(1)2}$ | 88.41(2) |
| $I^{(2)1}$ | Pb | $I^{(4)1}$ | 87.50(16) | $I^{(3A)3}$ | $Pb^{(1)}$ | $I^{(1)}$ | 91.59(2) |
| $I^{(2)1}$ | Pb | $I^{(4)}$ | 92.51(16) | $I^{(3A)3}$ | $Pb^{(1)}$ | $I^{(3A)4}$ | 180.0 |
| $I^{(2)}$ | Pb | $I^{(4)}$ | 87.49(16) | $Pb^{(1)}$ | $I^{(2)}$ | Pb | 152.5(4) |
| $I^{(2)}$ | Pb | $I^{(4)1}$ | 92.50(16) | Pb | $I^{(2A)}$ | $Pb^{(1)}$ | 151.99(3) |
| $I^{(2A)1}$ | Pb | $I^{(2A)}$ | 180.0 | $Pb^{(1)5}$ | $I^{(3)}$ | Pb | 152.5(3) |
| $I^{(2A)1}$ | Pb | $I^{(3A)}$ | 90.55(2) | Pb | $I^{(3A)}$ | $Pb^{(1)5}$ | 153.14(3) |
| $I^{(2A)1}$ | Pb | $I^{(3A)1}$ | 89.45(2) | $C^{(1)}$ | $N^{(1)}$ | $C^{(7)}$ | 110(2) |
| $I^{(2A)}$ | Pb | $I^{(3A)1}$ | 90.55(2) | $C^{(1A)}$ | $N^{(1A)}$ | $C^{(7A)}$ | 110(2) |
| $I^{(2A)}$ | Pb | $I^{(3A)}$ | 89.45(2) | $C^{(2)}$ | $C^{(1)}$ | $N^{(1)}$ | 122(2) |
| $I^{(2A)}$ | Pb | $I^{(4)1}$ | 90.74(2) | $C^{(2)}$ | $C^{(1)}$ | $C^{(6)}$ | 116(2) |
| $I^{(2A)1}$ | Pb | $I^{(4)}$ | 90.74(2) | $C^{(6)}$ | $C^{(1)}$ | $N^{(1)}$ | 121(2) |
| $I^{(2A)}$ | Pb | $I^{(4)}$ | 89.26(2) | $C^{(2)}$ | $C^{(1A)}$ | $N^{(1A)}$ | 121.1(19) |
| $I^{(2A)1}$ | Pb | $I^{(4)1}$ | 89.26(2) | $C^{(2)}$ | $C^{(1A)}$ | $C^{(6A)}$ | 114(2) |
| $I^{(3)1}$ | Pb | $I^{(3)}$ | 180.0 | $C^{(6A)}$ | $C^{(1A)}$ | $N^{(1A)}$ | 125(2) |
| $I^{(3)1}$ | Pb | $I^{(4)1}$ | 90.85(15) | $C^{(1)}$ | $C^{(2)}$ | $C^{(3)}$ | 127(2) |
| $I^{(3)}$ | Pb | $I^{(4)}$ | 90.85(15) | $C^{(3A)}$ | $C^{(2)}$ | $C^{(1A)}$ | 127.7(19) |
| $I^{(3)}$ | Pb | $I^{(4)1}$ | 89.15(15) | $C^{(4)}$ | $C^{(3)}$ | $C^{(2)}$ | 116(2) |
| $I^{(3)1}$ | Pb | $I^{(4)}$ | 89.15(15) | $C^{(2)}$ | $C^{(3A)}$ | $C^{(4A)}$ | 118(2) |
| $I^{(3A)1}$ | Pb | $I^{(3A)}$ | 180.0 | $C^{(5)}$ | $C^{(4)}$ | $C^{(3)}$ | 122(3) |

| Atom | Atom | Atom | Angle/° | Atom | Atom | Atom | Angle/° |
|---|---|---|---|---|---|---|---|
| $I^{(3A)1}$ | Pb | $I^{(4)1}$ | 91.66(2) | $C^{(5A)}$ | $C^{(4A)}$ | $C^{(3A)}$ | 120(3) |
| $I^{(3A)}$ | Pb | $I^{(4)}$ | 91.66(2) | $C^{(4)}$ | $C^{(5)}$ | $C^{(6)}$ | 122(3) |
| $I^{(3A)}$ | Pb | $I^{(4)1}$ | 88.34(2) | $C^{(4A)}$ | $C^{(5A)}$ | $C^{(6A)}$ | 124(3) |
| $I^{(3A)1}$ | Pb | $I^{(4)}$ | 88.34(2) | $C^{(1)}$ | $C^{(6)}$ | $C^{(5)}$ | 117(2) |
| $I^{(4)}$ | Pb | $I^{(4)1}$ | 180.0 | $C^{(1A)}$ | $C^{(6A)}$ | $C^{(5A)}$ | 118(2) |
| $I^{(1)}$ | $Pb^{(1)}$ | $I^{(1)2}$ | 180.0 | $C^{(8)}$ | $C^{(7)}$ | $N^{(1)}$ | 108(3) |
| $I^{(2)2}$ | $Pb^{(1)}$ | $I^{(1)}$ | 92.44(16) | $C^{(8A)}$ | $C^{(7A)}$ | $N^{(1A)}$ | 110(3) |
| $I^{(2)}$ | $Pb^{(1)}$ | $I^{(1)}$ | 87.56(16) | $C^{(9)}$ | $N^{(2)}$ | $C^{(15)}$ | 112.5(19) |
| $I^{(2)}$ | $Pb^{(1)}$ | $I^{(1)2}$ | 92.44(16) | $C^{(15)}$ | $N^{(2A)}$ | $C^{(9A)}$ | 111(2) |
| $I^{(2)2}$ | $Pb^{(1)}$ | $I^{(1)2}$ | 87.56(16) | $C^{(10)}$ | $C^{(9)}$ | $N^{(2)}$ | 121(3) |
| $I^{(2)}$ | $Pb^{(1)}$ | $I^{(2)2}$ | 180.0 | $C^{(14)}$ | $C^{(9)}$ | $N^{(2)}$ | 119(3) |
| $I^{(2)}$ | $Pb^{(1)}$ | $I^{(3)3}$ | 90.0(2) | $C^{(14)}$ | $C^{(9)}$ | $C^{(10)}$ | 120(3) |
| $I^{(2)}$ | $Pb^{(1)}$ | $I^{(3)4}$ | 90.0(2) | $C^{(10)}$ | $C^{(9A)}$ | $N^{(2A)}$ | 121(3) |
| $I^{(2)2}$ | $Pb^{(1)}$ | $I^{(3)4}$ | 90.0(2) | $C^{(14A)}$ | $C^{(9A)}$ | $N^{(2A)}$ | 118(3) |
| $I^{(2)2}$ | $Pb^{(1)}$ | $I^{(3)3}$ | 90.0(2) | $C^{(14A)}$ | $C^{(9A)}$ | $C^{(10)}$ | 120(3) |
| $I^{(2A)2}$ | $Pb^{(1)}$ | $I^{(1)2}$ | 89.29(2) | $C^{(11)}$ | $C^{(10)}$ | $C^{(9)}$ | 113(2) |
| $I^{(2A)}$ | $Pb^{(1)}$ | $I^{(1)2}$ | 90.71(2) | $C^{(11A)}$ | $C^{(10)}$ | $C^{(9A)}$ | 116(2) |
| $I^{(2A)}$ | $Pb^{(1)}$ | $I^{(1)}$ | 89.29(2) | $C^{(10)}$ | $C^{(11)}$ | $C^{(12)}$ | 128(3) |
| $I^{(2A)2}$ | $Pb^{(1)}$ | $I^{(1)}$ | 90.71(2) | $C^{(10)}$ | $C^{(11A)}$ | $C^{(12A)}$ | 123(3) |
| $I^{(2A)}$ | $Pb^{(1)}$ | $I^{(2A)2}$ | 180.0 | $C^{(13)}$ | $C^{(12)}$ | $C^{(11)}$ | 114(3) |
| $I^{(2A)}$ | $Pb^{(1)}$ | $I^{(3A)4}$ | 90.57(2) | $C^{(13A)}$ | $C^{(12A)}$ | $C^{(11A)}$ | 119(3) |
| $I^{(2A)}$ | $Pb^{(1)}$ | $I^{(3A)3}$ | 89.43(2) | $C^{(14)}$ | $C^{(13)}$ | $C^{(12)}$ | 122(3) |
| $I^{(2A)2}$ | $Pb^{(1)}$ | $I^{(3A)4}$ | 89.43(2) | $C^{(14A)}$ | $C^{(13A)}$ | $C^{(12A)}$ | 118(3) |
| $I^{(2A)2}$ | $Pb^{(1)}$ | $I^{(3A)3}$ | 90.57(2) | $C^{(13)}$ | $C^{(14)}$ | $C^{(9)}$ | 124(3) |
| $I^{(3)3}$ | $Pb^{(1)}$ | $I^{(1)2}$ | 89.17(16) | $C^{(13A)}$ | $C^{(14A)}$ | $C^{(9A)}$ | 124(3) |
| $I^{(3)4}$ | $Pb^{(1)}$ | $I^{(1)2}$ | 90.83(16) | $C^{(16)}$ | $C^{(15)}$ | $N^{(2)}$ | 112.0(14) |
| $I^{(3)3}$ | $Pb^{(1)}$ | $I^{(1)}$ | 90.83(16) | $C^{(16)}$ | $C^{(15)}$ | $N^{(2A)}$ | 112.9(14) |

$^{1}$-X,1-Y,1-Z; $^{2}$1-X,2-Y,1-Z; $^{3}$1-X,1-Y,1-Z; $^{4}$+X,1+Y,+Z; $^{5}$+X,-1+Y,+Z

Table S 6: Torsion Angles for $(PEA)_2PbI_4$.

| A | B | C | D | Angle/° | A | B | C | D | Angle/° |
|---|---|---|---|---|---|---|---|---|---|
| $N^{(1)}$ | $C^{(1)}$ | $C^{(2)}$ | $C^{(3)}$ | -180(3) | $N^{(2)}$ | $C^{(9)}$ | $C^{(10)}$ | $C^{(11)}$ | 177(2) |
| $N^{(1)}$ | $C^{(1)}$ | $C^{(6)}$ | $C^{(5)}$ | 180(3) | $N^{(2)}$ | $C^{(9)}$ | $C^{(14)}$ | $C^{(13)}$ | -178(3) |
| $N^{(1A)}$ | $C^{(1A)}$ | $C^{(2)}$ | $C^{(3A)}$ | 178(3) | $N^{(2A)}$ | $C^{(9A)}$ | $C^{(10)}$ | $C^{(11A)}$ | -175(3) |
| $N^{(1A)}$ | $C^{(1A)}$ | $C^{(6A)}$ | $C^{(5A)}$ | -180(3) | $N^{(2A)}$ | $C^{(9A)}$ | $C^{(14A)}$ | $C^{(13A)}$ | 179(3) |
| $C^{(1)}$ | $N^{(1)}$ | $C^{(7)}$ | $C^{(8)}$ | -61(3) | $C^{(9)}$ | $N^{(2)}$ | $C^{(15)}$ | $C^{(16)}$ | 60(3) |
| $C^{(1)}$ | $C^{(2)}$ | $C^{(3)}$ | $C^{(4)}$ | 4(4) | $C^{(9)}$ | $C^{(10)}$ | $C^{(11)}$ | $C^{(12)}$ | 0(4) |

| A | B | C | D | Angle/° | A | B | C | D | Angle/° |
|---|---|---|---|---|---|---|---|---|---|
| $C^{(1A)}$ | $N^{(1A)}$ | $C^{(7A)}$ | $C^{(8A)}$ | 63(3) | $C^{(9A)}$ | $N^{(2A)}$ | $C^{(15)}$ | $C^{(16)}$ | -62(2) |
| $C^{(1A)}$ | $C^{(2)}$ | $C^{(3A)}$ | $C^{(4A)}$ | 1(5) | $C^{(9A)}$ | $C^{(10)}$ | $C^{(11A)}$ | $C^{(12A)}$ | -5(4) |
| $C^{(2)}$ | $C^{(1)}$ | $C^{(6)}$ | $C^{(5)}$ | 3(4) | $C^{(10)}$ | $C^{(9)}$ | $C^{(14)}$ | $C^{(13)}$ | -3(4) |
| $C^{(2)}$ | $C^{(1A)}$ | $C^{(6A)}$ | $C^{(5A)}$ | -2(4) | $C^{(10)}$ | $C^{(9A)}$ | $C^{(14A)}$ | $C^{(13A)}$ | 3(5) |
| $C^{(2)}$ | $C^{(3)}$ | $C^{(4)}$ | $C^{(5)}$ | -5(5) | $C^{(10)}$ | $C^{(11)}$ | $C^{(12)}$ | $C^{(13)}$ | 0(5) |
| $C^{(2)}$ | $C^{(3A)}$ | $C^{(4A)}$ | $C^{(5A)}$ | -1(6) | $C^{(10)}$ | $C^{(11A)}$ | $C^{(12A)}$ | $C^{(13A)}$ | 7(5) |
| $C^{(3)}$ | $C^{(4)}$ | $C^{(5)}$ | $C^{(6)}$ | 5(5) | $C^{(11)}$ | $C^{(12)}$ | $C^{(13)}$ | $C^{(14)}$ | -1(5) |
| $C^{(3A)}$ | $C^{(4A)}$ | $C^{(5A)}$ | $C^{(6A)}$ | -1(6) | $C^{(11A)}$ | $C^{(12A)}$ | $C^{(13A)}$ | $C^{(14A)}$ | -4(5) |
| $C^{(4)}$ | $C^{(5)}$ | $C^{(6)}$ | $C^{(1)}$ | -5(5) | $C^{(12)}$ | $C^{(13)}$ | $C^{(14)}$ | $C^{(9)}$ | 3(5) |
| $C^{(4A)}$ | $C^{(5A)}$ | $C^{(6A)}$ | $C^{(1A)}$ | 3(5) | $C^{(12A)}$ | $C^{(13A)}$ | $C^{(14A)}$ | $C^{(9A)}$ | -1(5) |
| $C^{(6)}$ | $C^{(1)}$ | $C^{(2)}$ | $C^{(3)}$ | -3(4) | $C^{(14)}$ | $C^{(9)}$ | $C^{(10)}$ | $C^{(11)}$ | 2(3) |
| $C^{(6A)}$ | $C^{(1A)}$ | $C^{(2)}$ | $C^{(3A)}$ | 0(4) | $C^{(14A)}$ | $C^{(9A)}$ | $C^{(10)}$ | $C^{(11A)}$ | 0(4) |
| $C^{(7)}$ | $N^{(1)}$ | $C^{(1)}$ | $C^{(2)}$ | -72(3) | $C^{(15)}$ | $N^{(2)}$ | $C^{(9)}$ | $C^{(10)}$ | -102(3) |
| $C^{(7)}$ | $N^{(1)}$ | $C^{(1)}$ | $C^{(6)}$ | 111(3) | $C^{(15)}$ | $N^{(2)}$ | $C^{(9)}$ | $C^{(14)}$ | 73(3) |
| $C^{(7A)}$ | $N^{(1A)}$ | $C^{(1A)}$ | $C^{(2)}$ | 73(3) | $C^{(15)}$ | $N^{(2A)}$ | $C^{(9A)}$ | $C^{(10)}$ | 100(3) |
| $C^{(7A)}$ | $N^{(1A)}$ | $C^{(1A)}$ | $C^{(6A)}$ | -109(3) | $C^{(15)}$ | $N^{(2A)}$ | $C^{(9A)}$ | $C^{(14A)}$ | -76(3) |

Table S 7: Hydrogen Atom Coordinates (Å×$10^4$) and Isotropic Displacement Parameters (U [Å$^2$×$10^3$]) for $(PEA)_2PbI_4$.

| Atom | *x* | *y* | *z* | U(eq) |
|---|---|---|---|---|
| $H^{(1A)}$ | 5569 | 2838 | 2075 | 54 |
| $H^{(1B)}$ | 3903 | 2491 | 1896 | 54 |
| $H^{(1AA)}$ | 5982 | 4582 | 1923 | 55 |
| $H^{(1AB)}$ | 5703 | 3009 | 2102 | 55 |
| $H^{(2)}$ | 2227 | 4759 | 1572 | 55 |
| $H^{(3)}$ | 1880 | 6731 | 811 | 66 |
| $H^{(3A)}$ | 1424 | 6192 | 761 | 69 |
| $H^{(4)}$ | 4022 | 7563 | 218 | 68 |
| $H^{(4A)}$ | 66 | 3841 | 225 | 88 |
| $H^{(5)}$ | 6360 | 6775 | 547 | 68 |
| $H^{(5A)}$ | 964 | 1617 | 521 | 73 |
| $H^{(6)}$ | 6764 | 4663 | 1340 | 49 |
| $H^{(6A)}$ | 3389 | 1477 | 1345 | 55 |
| $H^{(7A)}$ | 3302 | 4453 | 3098 | 55 |
| $H^{(7B)}$ | 4406 | 3164 | 3406 | 55 |
| $H^{(7AA)}$ | 6034 | 4839 | 3355 | 54 |
| $H^{(7AB)}$ | 4563 | 5728 | 3032 | 54 |
| $H^{(8A)}$ | 6545 | 4734 | 3375 | 69 |
| $H^{(8B)}$ | 5350 | 5688 | 3950 | 69 |

| Atom | *x* | *y* | *z* | U(eq) |
|---|---|---|---|---|
| $H^{(8C)}$ | 5477 | 6004 | 3031 | 69 |
| $H^{(8AA)}$ | 4170 | 2537 | 3150 | 68 |
| $H^{(8AB)}$ | 2896 | 3860 | 3228 | 68 |
| $H^{(8AC)}$ | 4173 | 3815 | 3979 | 68 |
| $H^{(2A)}$ | 83 | -2465 | 1941 | 59 |
| $H^{(2B)}$ | -1526 | -2122 | 2105 | 59 |
| $H^{(2AA)}$ | 952 | -1674 | 1955 | 56 |
| $H^{(2AB)}$ | 719 | 25 | 2110 | 56 |
| $H^{(10)}$ | -2982 | -190 | 1446 | 60 |
| $H^{(11)}$ | -2727 | 1734 | 626 | 69 |
| $H^{(11A)}$ | -3969 | 637 | 711 | 65 |
| $H^{(12)}$ | -365 | 2612 | 241 | 78 |
| $H^{(12A)}$ | -4861 | -1951 | 197 | 66 |
| $H^{(13)}$ | 1869 | 1456 | 731 | 66 |
| $H^{(13A)}$ | -3545 | -4039 | 699 | 66 |
| $H^{(14)}$ | 1732 | -344 | 1564 | 52 |
| $H^{(14A)}$ | -1334 | -3469 | 1528 | 50 |
| $H^{(15A)}$ | -11 | -1810 | 3367 | 49 |
| $H^{(15B)}$ | 1022 | -538 | 3056 | 49 |
| $H^{(16A)}$ | -836 | 1184 | 3151 | 33 |
| $H^{(16B)}$ | -854 | 446 | 3973 | 33 |
| $H^{(16C)}$ | -2117 | -109 | 3213 | 33 |

Table S 8: Atomic Occupancy for $(PEA)_2PbI_4$.

| Atom | *Occupancy* | Atom | *Occupancy* | Atom | *Occupancy* |
|---|---|---|---|---|---|
| $I^{(2)}$ | 0.122(3) | $I^{(2A)}$ | 0.878(3) | $I^{(3)}$ | 0.121(2) |
| $I^{(3A)}$ | 0.879(2) | $N^{(1)}$ | 0.5 | $H^{(1A)}$ | 0.5 |
| $H^{(1B)}$ | 0.5 | $N^{(1A)}$ | 0.5 | $H^{(1AA)}$ | 0.5 |
| $H^{(1AB)}$ | 0.5 | $C^{(1)}$ | 0.5 | $C^{(1A)}$ | 0.5 |
| $C^{(3)}$ | 0.5 | $H^{(3)}$ | 0.5 | $C^{(3A)}$ | 0.5 |
| $H^{(3A)}$ | 0.5 | $C^{(4)}$ | 0.5 | $H^{(4)}$ | 0.5 |
| $C^{(4A)}$ | 0.5 | $H^{(4A)}$ | 0.5 | $C^{(5)}$ | 0.5 |
| $H^{(5)}$ | 0.5 | $C^{(5A)}$ | 0.5 | $H^{(5A)}$ | 0.5 |
| $C^{(6)}$ | 0.5 | $H^{(6)}$ | 0.5 | $C^{(6A)}$ | 0.5 |
| $H^{(6A)}$ | 0.5 | $C^{(7)}$ | 0.5 | $H^{(7A)}$ | 0.5 |
| $H^{(7B)}$ | 0.5 | $C^{(7A)}$ | 0.5 | $H^{(7AA)}$ | 0.5 |
| $H^{(7AB)}$ | 0.5 | $C^{(8)}$ | 0.5 | $H^{(8A)}$ | 0.5 |

| **Atom** | ***Occupancy*** | **Atom** | ***Occupancy*** | **Atom** | ***Occupancy*** |
|---|---|---|---|---|---|
| $H^{(8B)}$ | 0.5 | $H^{(8C)}$ | 0.5 | $C^{(8A)}$ | 0.5 |
| $H^{(8AA)}$ | 0.5 | $H^{(8AB)}$ | 0.5 | $H^{(8AC)}$ | 0.5 |
| $N^{(2)}$ | 0.508(9) | $H^{(2A)}$ | 0.508(9) | $H^{(2B)}$ | 0.508(9) |
| $N^{(2A)}$ | 0.492(9) | $H^{(2AA)}$ | 0.492(9) | $H^{(2AB)}$ | 0.492(9) |
| $C^{(9)}$ | 0.508(9) | $C^{(9A)}$ | 0.492(9) | $C^{(11)}$ | 0.508(9) |
| $H^{(11)}$ | 0.508(9) | $C^{(11A)}$ | 0.492(9) | $H^{(11A)}$ | 0.492(9) |
| $C^{(12)}$ | 0.508(9) | $H^{(12)}$ | 0.508(9) | $C^{(12A)}$ | 0.492(9) |
| $H^{(12A)}$ | 0.492(9) | $C^{(13)}$ | 0.508(9) | $H^{(13)}$ | 0.508(9) |
| $C^{(13A)}$ | 0.492(9) | $H^{(13A)}$ | 0.492(9) | $C^{(14)}$ | 0.508(9) |
| $H^{(14)}$ | 0.508(9) | $C^{(14A)}$ | 0.492(9) | $H^{(14A)}$ | 0.492(9) |

Clarification of the difference between phases 1 (or 3) and 2 (or 4)

Each configuration can be divided into an inorganic part (the $PbI_6$ octahedra) and an organic part (PEA molecules). Phases 1 and 2 form chiral images of only the inorganic part where each octahedron is rotated around the *c** axis. The difference in band gap and total energy comes from differences in the PEA molecule attachment to the inorganic complex (figure S2). Phases 3 and 4 are different from phases 1 and 2 by the orientation of the organic part. Each PEA molecule is rotated around an axis (not the *c** axis). This means that there is a slight structural difference between all four phases. The rotation axis of the PEA molecule is determined because the relaxed N atom is fixed in position with respect to phases 1 (or 3) and 2 (or 4).

Hence, the rotation axis must go through it (this also excludes the possibility of rotation and translation of the PEA molecule). See Figure S2 for visual clarification.

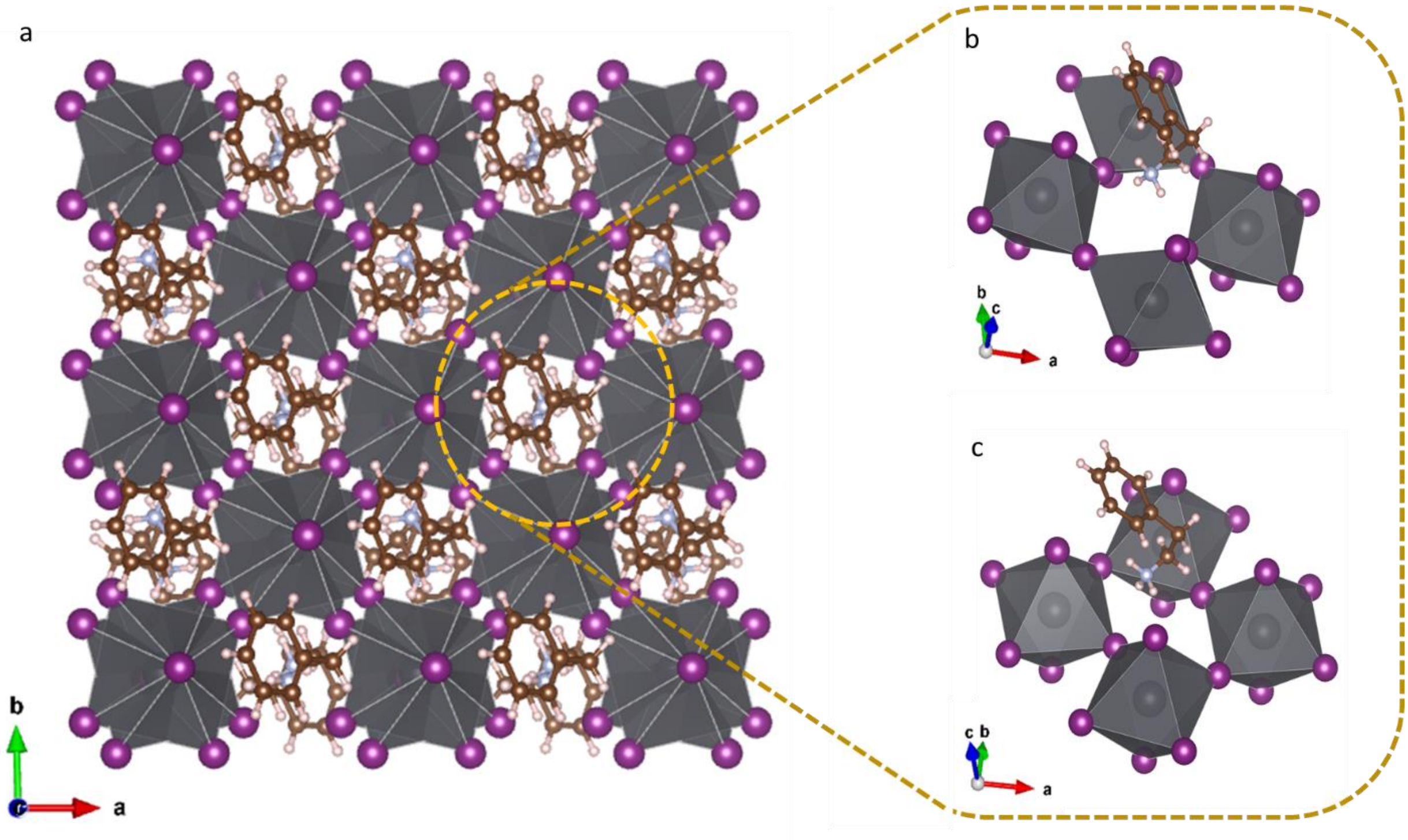

**Figure S2: a. Supper imposition of phases 1 and 2 with magnification on the local environment of the amine group. b. Amine group attached to phase 1. c. Amine group attached to phase 2.**

## Power dependence photoluminescence (PL)

To exclude the formation of biexcitons as the source of the dual band edge emission, Power dependent measurements were made. We used a 488 nm laser to excite the samples at different excitation powers. Fig. S2a shows the power dependence spectra close to the band edge energy, where the two components at 2.342 and 2.349 eV are the surface and bulk emissions, respectively (as explained in the main text). The extracted PL spectra were fitted with a three gaussian model to account for the tail emission at lower energies and was used to calculate the integrated intensities of the bulk and surface emissions in Fig. S2b.

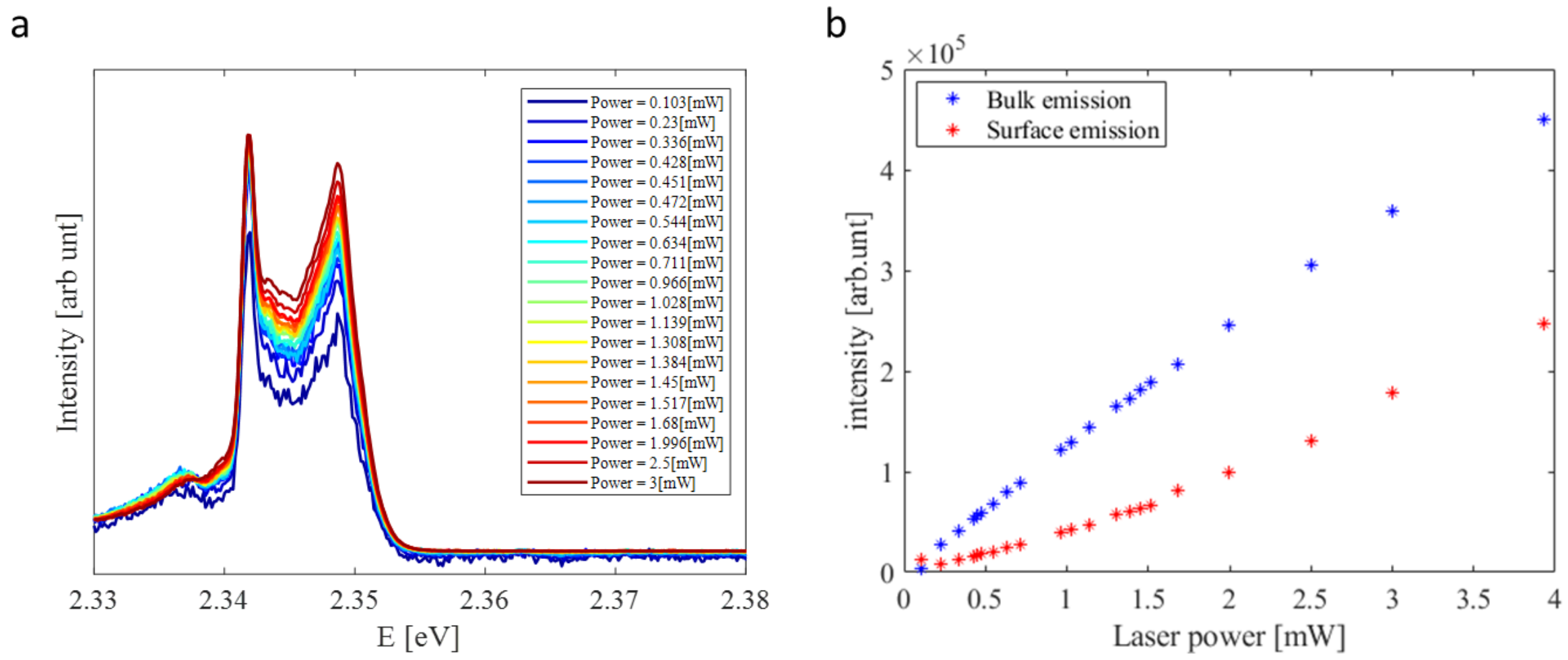

**Figure S3: Power dependent PL. a.** PL excited by a 488 nm laser, as a function of excitation power. **b.** Power-dependent integrated intensity, extracted from the PL spectra in panel a, calculated by fitting a 3 gaussian model where the third gaussian models the peak at 2.336 eV.

## Temperature dependence photoluminescence (TPL)

To emphasize the temperature dependence of the PL, we show the TPL at individual temperatures in figure S4. The intensity was normalized in respect to the maximum of each individual spectrum. We observe that the dual emissions persist to higher temperatures after the expected phase transition around ~20 K accompanied by a broadening of the peaks. Due to a similarity in the band gaps of phases 1 and 2 compared to phases 4 and 3, no further peaks are observed above the phase transition temperature.

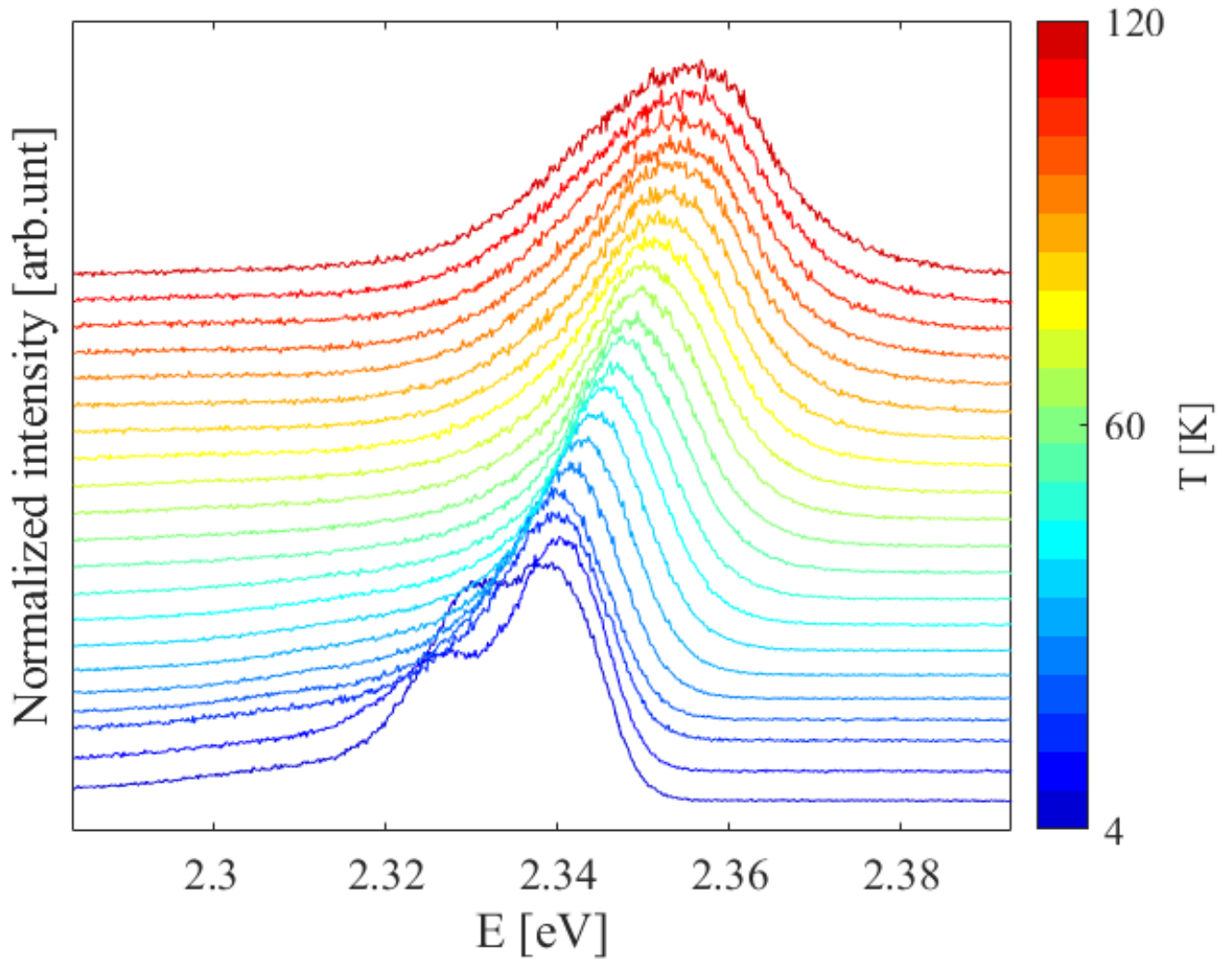

Figure S4: TPL plot of $(PEA)_2PbI_4$ single crystal. The intensity was normalized in respect to the maximum emission of the individual spectrum.

## Computational Details

To determine the ground state of $(PEA)_2PbI_4$ we conducted DFT calculations as detailed in the method section. The assessment of the ground state was based on the fully relaxed total DFT energies structures of phases 1-4 and are summarized in Tab. S9.

**Table S 9: Fully relaxed PBE+SOC+ D3 total energies for configurations 1-4.**

| conf | [Ry/cell] | [Ry/atom] | [eV/cell] | [eV/atom] | dE [meV/atom] |
|---|---|---|---|---|---|
| 1 | -1014.225 | -10.789 | -13799.24 | -146.80 | |
| 2 | -1014.226 | -10.789 | -13799.25 | -146.80 | -0.1 |
| 3 | -1014.212 | -10.789 | -13799.07 | -146.79 | 1.8 |
| 4 | -1014.212 | -10.789 | -13799.07 | -146.795 | 1.9 |

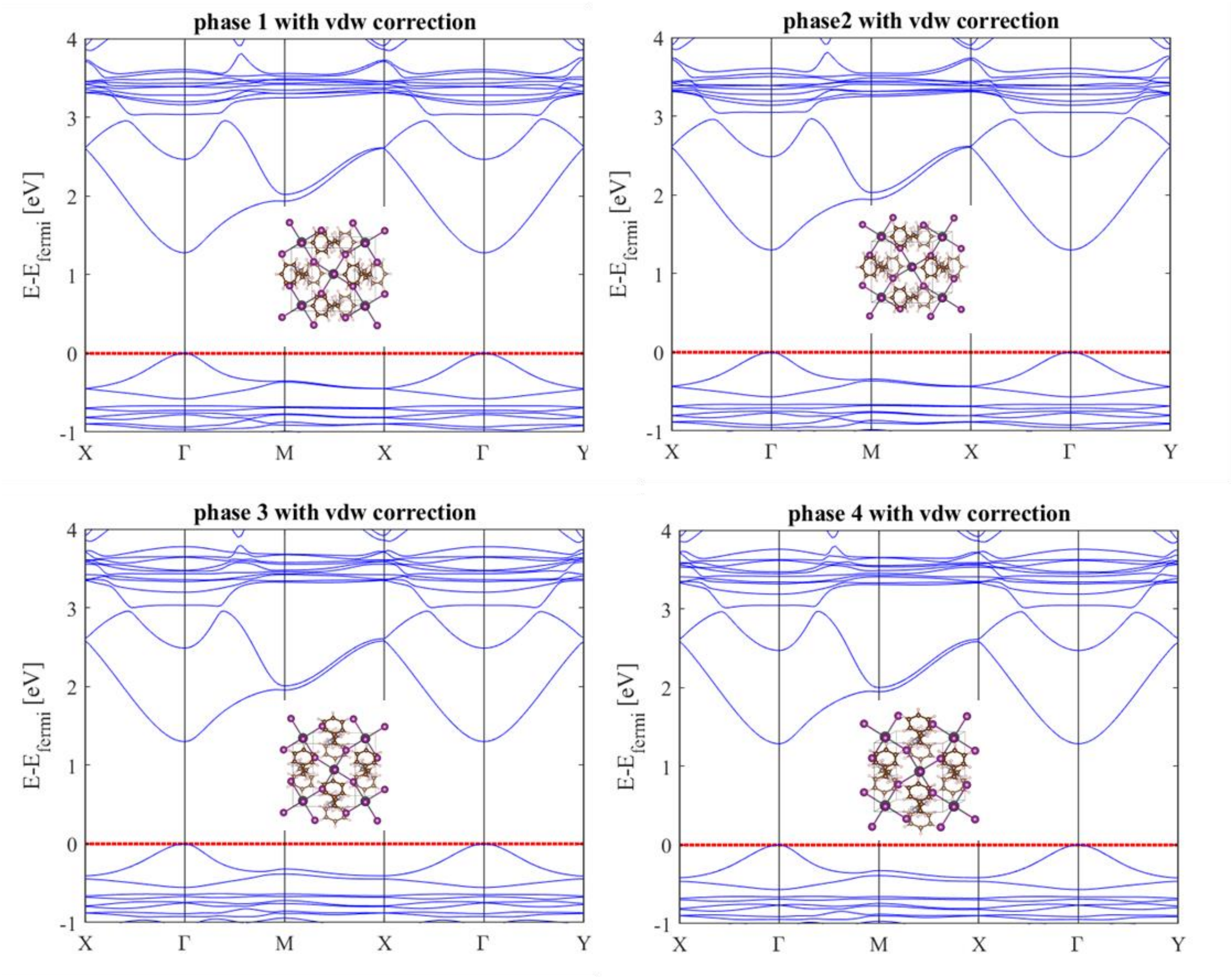

**Figure S5: DFT band structure of configurations 1-4.**